          [2001/04/25 1.1 (PWD)]
\documentclass[a4paper,twocolumn]{esapub} 
\usepackage{natbib}

\usepackage{psfig}
\usepackage{epsfig}
\usepackage{graphicx}

\title{The first broad-band persistent X-ray spectrum of 
the dipping 
low mass X-ray binary  EXO 0748-676}
\author{L. Sidoli}
\affil{IASF, 20133 Milano,
Italy}
\author{A. N. Parmar}
\affil{RSSD, ESTEC, 2200 AG Noordwijk, The Netherlands}
\author{T. Oosterbroek}
\affil{INTEGRAL SOC, SODSD, RSSD, ESTEC, 2200 AG Noordwijk, The Netherlands}

\begin{document}


\maketitle

\begin{abstract}
We report on the results of a BeppoSAX observation of the dipping 
LMXRB EXO 0748-676 performed in 2000 November. 
This is the first simultaneous observation of this source over the 0.1-100 keV 
energy range. The persistent spectrum is complex and shows a soft excess, 
which requires the inclusion of an ionized absorber (with a sub-solar 
abundance of iron). 
A cutoff power-law is a good fit to the high energy part of the 
spectrum, with a photon index of 1.3 and a cutoff around 50 keV. 
The 0.1-100 keV luminosity is 8.2$\times$10$^{36}$ erg~s$^{-1}$.
\end{abstract}

\section{Introduction}

EXO 0748-676 is a transient LMXRB discovered with EXOSAT 
(Parmar et al., 1986) 
with a 3.82~hr orbital period measured from its X-ray eclipses. 
It is a type I X-ray burster and displays quasi periodic oscillations 
(discovered with RossiXTE, Homan et al., 1999; 
Homan \& van der Klis, 2000) as well as intensity 
dips and X--ray eclipses (Parmar et al. 1986; Church et al., 1998). 

It shows a complex soft X-ray spectrum (Thomas et al., 1997). 
Church et al. (1998), analysing dip and non-dip ASCA spectra, 
interpreted the source spectrum as composed by an  
extended Accretion Disk Corona (ADC), producing the 
hard emission, plus a point-like soft 
black-body emission. The spectral evolution during dipping 
is explained with the  progressive covering of the
ADC emission. 

A different view on the system came from EPIC/XMM-Newton
 observations, which 
revealed a persistent (i.e., no dipping and no eclipsing) 
spectrum composed by an extended  thermal halo together with a highly
 absorbed, and more compact, high energy power-law produced in the ADC
 (Bonnet-Bidaud et al., 2001).

RGS/XMM-Newton observations 
detected absorption and emission lines from ionized Ne, O and N 
(Cottam et al.,2001). 

The cumulative RGS spectra during type I X-ray bursts revealed 
iron absorption lines, from which a gravitational redshift of z=0.35 
have been measured (Cottam et al., 2002).

Chandra observations confirmed the presence of photoionized plasma, 
probably located above the accretion disk. 
The Chandra spectrum 
during dips was absorbed both by neutral and ionized material 
(Jimenez-Garate et al., 2003). 

A recent analysis of several XMM-Newton observations revealed for the first
time the clear detection of eclipses below 2 keV (Homan et al. 2003).

\section{BeppoSAX Observations}

The source was observed with BeppoSAX in 2000, November, 
for an on-source time of 66~ks. 
We report here spectral results from all the instruments on-board
BeppoSAX: 
LECS (0.1-10 keV), MECS (1.8-10.8 keV), HPGSPC (5-120 keV) 
and PDS (15-200 keV) (Sidoli et al., 2004).

The source lightcurve in the 1.8-10 keV band is shown in Fig.~1, where all kind
of variabilities typical of this source are evident: 
X--ray eclipses, dips, type I X--ray
bursts.


\begin{small}
\begin{figure*}
\centering
\includegraphics[width=7cm,angle=-90]{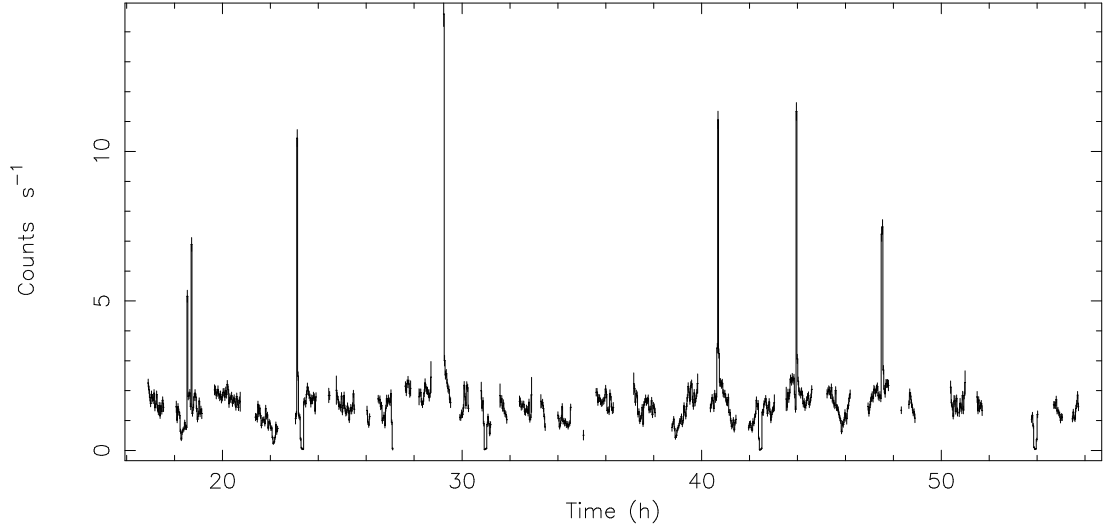}
\caption{BeppoSAX MECS lightcurve 
(1.8-10 keV) with a time binning of 128s.  
Type I X-ray bursts, dips, and eclipses are clearly evident
}
\end{figure*}
\end{small}


\section{Spectral Results}

We concentrate here on the ``persistent" spectrum 
(no-dipping, no-eclipsing, no-bursting spectrum). 
In order to select the persistent emission, we considered
only events where the source displays
a constant low hardness ratio (see Fig.~2).

\begin{small}
\begin{figure}
\centering
\includegraphics[width=6cm,angle=-90]{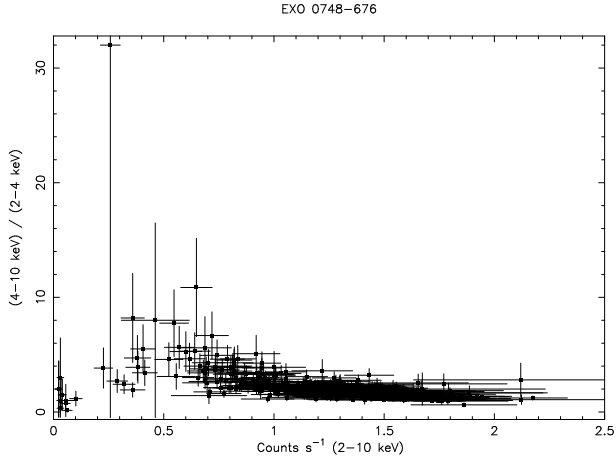}
\caption{EXO~0748-676 hardness ratio versus 2--10 keV intensity.
Time intervals containing bursts have been excluded. 
In order to consider only events with a low  hardness ratio
we selected 2--10~keV intensity $>$1.5~s$^{-1}$
}
\end{figure}
\end{small}


\begin{small}
\begin{figure}
\centering
\includegraphics[width=5.0cm,angle=-90]{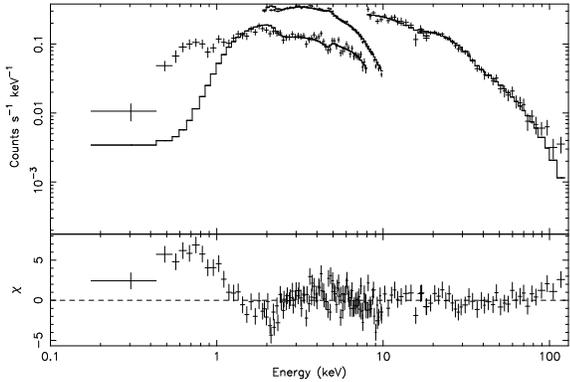}
\caption{The fit to the BeppoSAX persistent spectrum  with an absorbed cutoff 
powerlaw clearly is inadequate and shows a soft excess. The lower panel
shows the residuals in units of standard deviations 
}
\end{figure}
\end{small}


We tried to fit the ``persistent" spectrum with several models, all including
a high energy cut-off powerlaw in order to account 
for the high energy part of the spectrum.
In Fig.~2  the count spectrum (together with the residuals) is reported, 
fit with an absorbed cut-off powerlaw. 
There is clearly evidence for a soft
excess and other structured residuals, especially 
in the low energy part of the spectrum.
Trying to model this soft excess with soft components (blackbody or hot
plasma models (i.e. {\sc mekal} in {\sc xpsec})) 
or different kinds
of absorptions (e.g. partial covering model, differential absorbing models for
the soft and the hard spectral components) resulted in non-satisfactory
fits ($\chi^2$ $>$2).  

Among all spectral deconvolutions tested, 
a cut-off power-law absorbed with an ionized absorber (with a non-solar
iron abundance) resulted 
in a significanlty better fit. 
An additional absorption line
at $\sim$2.16~keV is needed to get a good fit to the spectrum (see Table~1 for the
best-fit parameters).

\begin{table}
\begin{center}
\caption[]{Best-fit parameters for the broad-band BeppoSAX
``persistent'' spectrum of EXO 0748-676.
The meaning of the symbols is the following: 
$N_{\rm H}$ is the interstellar column density, $N_{\rm absori}$ is
the local absorbing column density due to the ionized
absorber. $\xi$ is the ionization parameter ($\xi$=L/nR$^{2}$, where
$L$ is the luminosity of the X--ray illuminating source, $n$ is the absorbing 
plasma density and $R$ is the distance of the absorbing matter from the 
ionizing source). The parameters of the cut-off power-law are
$\Gamma$, the cut-off power-law photon index and
$E_{\rm c}$, the high energy cut-off. 
$E_{\rm line}$ is the  centroid energy of the absorption 
line with a width $\sigma$ and an intensity $I_{\rm line}$.
The fluxes reported here are ``observed" fluxes (not corrected for the absorption). 
The luminosity  
has been corrected {\it only} for interstellar absorption and calculated for a 
distance of 10~kpc
}
\begin{tabular}{ll}
\hline
\noalign {\smallskip}
Parameter & Value   \\
\hline
\noalign {\smallskip}
$N_{\rm H}$ $(10^{22}$ cm$^{-2}$)  		&  $0.12  ^{+0.03} _{-0.02} $     \\
$N_{\rm absori}$ $(10^{22}$ cm$^{-2}$) 		 &  $5  ^{+3} _{-1} $     \\
$\xi$                                  		 & $ 250 \pm{60} $   \\
Fe abundance                   			&  $0.6 ^{+0.3} _{-0.4}$   \\
$\Gamma$                      			 &   $1.33 ^{+0.08} _{-0.05}$      \\
$E_{\rm c}$          (keV)             		 &    $47 \pm{7}$    \\
$E_{\rm line}$   (keV)    			 &    $ 2.16 ^{+0.05} _{-0.04}$    \\
$I_{\rm line}$  (${\rm  10^{-4} photons~cm^{-2}~s^{-1}})$      &    $-4.2 ^{+1.5} _{-2.0}$    \\
$EW$ (eV)							&    $ -40 ^{+15} _{-20}$    \\
Flux (2--10 keV) (${\rm erg~cm^{-2}~s^{-1}})$      		 &  $ 1.7\times 10^{-10}$    \\
Flux (0.1--100 keV) (${\rm erg~cm^{-2}~s^{-1}})$       		&  $  6.9 \times 10^{-10}$    \\
Luminosity (2--10 keV) (${\rm erg~s^{-1}})$          		&   $ 2.0\times 10^{36}$    \\
Luminosity (0.1--100 keV) (${\rm erg~s^{-1}})$          		&   $ 8.2\times 10^{36}$    \\
$\chi ^2$/dof                            			&  189.6/168            \\
\noalign {\smallskip}
\hline
\label{tab:spec}
\end{tabular}
\end{center}
\end{table}

\begin{small}
\begin{figure}
\centering
\includegraphics[width=6.9cm,angle=-90]{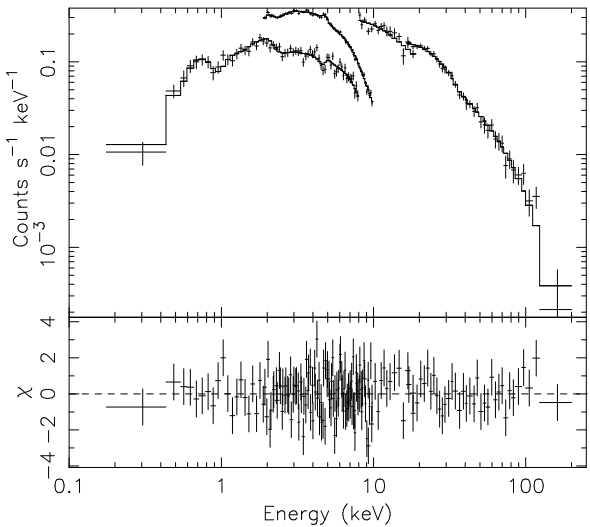}
\caption{BeppoSAX best-fit: cut-off powerlaw with an ionized absorber.
See Table~1 for the spectral parameters}
\end{figure}
\end{small}


Currently a reanalysis of the data is in progress taking into account
the recent results of Homan et al. (2003) who found a change in the
spectrum below 2 keV, which is not visible above 2 keV (i.e. our
selection criterion). This represents low-level dipping activity
and might cause ``mixing" of spectra which might affect our results
at low energies.

\vskip -1.5cm
\section{Conclusions}

We have discussed here the first  observation of the
broad-band X--ray emission of the dipping LMXRB EXO~0748-676
showing that this source displays significant emission up to 100~keV.

The best-fit model to the ``persistent" spectrum consists of 
a cut-off power-law, extra-absorbed at low energies by
an ionized absorber with non-solar iron abundance.
The need for an ionized absorber  indicates that
absorbing matter local to the binary system is located over
most or all of the orbital period.


\end{document}